\documentclass[aps,twocolumn,showpacs]{revtex4}
\usepackage{graphicx}

\begin{document}

\title{Carbon cage-like materials as potential
low work function metallic compounds: \\
Case of clathrates.}

\author{V. Timoshevskii}
\email{vladimir@dpm.univ-lyon1.fr}
\author{D. Conn\'etable}
\author{X. Blase}
\affiliation{D\'epartement de Physique des Mat\'eriaux and CNRS,\\
Universit\'e Claude Bernard-Lyon 1, B\^at. 203, 43 Bd du 11
Novembre 1918, \\ 69622 Villeurbanne C\'edex, FRANCE }

\begin{abstract}
We present an {\it ab-initio} calculation of the electronic affinity of the
hypothetical C-46 clathrate by studying its bare and hydrogenated (100)
surfaces. We show that such  a system shares with the diamond phase
a small electronic affinity. Further, contrary to the diamond phase,
the possibility of doping endohedrally these cage-like systems allows
to significantly raise the position of the Fermi level, resulting in a true
metal with a small work function. This is illustrated in the case of the
Li$_8$@C-46  doped compound. Such a class of materials might be of much
interest for the design of electron-emitting devices.
\end{abstract}


\maketitle


There is a strong incentive in trying to synthesize materials which
could be used as efficient electron-emitters for vacuum electronic devices.
While alkali metals display the lowest work function (W$\sim$2-3 eV) of all
elemental metals, their poor mechanical and chemical properties (melting
temperature, resistance to ionic bombardment, etc) forbid their use in actual
devices, and therefore metals with much higher work function such as tungsten
or molybdinum (W$\sim$4.5 eV) are preferred for their superior mechanical
and chemical properties.

A promising class of materials which have been extensively studied
are carbon diamond based systems. Due to their large gap,
carbon diamond based
systems have been shown to display a small, or even negative,
electronic affinity \cite{Weide94,Pickett94}. Further, their mechanical
properties are superior to the one of any existing metal. However,
the difficulty of doping them "n-type" in order to obtain a significant density
of emitting electrons limits their efficiency as good emitters.


In the last few years, a novel structure has attracted much
attention for its specific structural and electronic properties.
Column-IV clathrates \cite{Kasper65} are cage-like materials
composed of face-sharing X$_{20}$, X$_{24}$ and X$_{28}$ clusters
(where X=Si,Ge). All atoms are in the same sp$^3$ environment as
in the diamond phase, but the network is composed of 87$\%$ of
pentagonal rings. As a result, it has been shown in particular
that Si clathrates present a band gap which is $\sim$ 0.7 eV
larger than the one of the diamond phase \cite{Adams94,Gryco00}. Furthermore,
the possibility of filling each cage by a dopant atom allows to
significantly raise the position of the Fermi level to yield a
metallic system with a large density of states (eDOS) at the Fermi
level (E$_F$) \cite{Saito95,Smelyansky97}.

The possibility of efficiently doping such  systems has incited us to study
the case of the hypothetical carbon clathrates.  Even though such phases
have not been synthesized so far, the reports on the synthesis
of connected C$_{36}$ clusters \cite{Piskoti98}, of diamond-like thin films
synthesized by deposition of small carbon  clusters \cite{Paillard93} or even
of polymerized C$_{60}$ suggests that the synthesis of such phases might be at
hand or that the results presented below may apply to the above-mentioned
existing phases.

\begin{figure}
\begin{center}
\includegraphics[width=8.5cm]{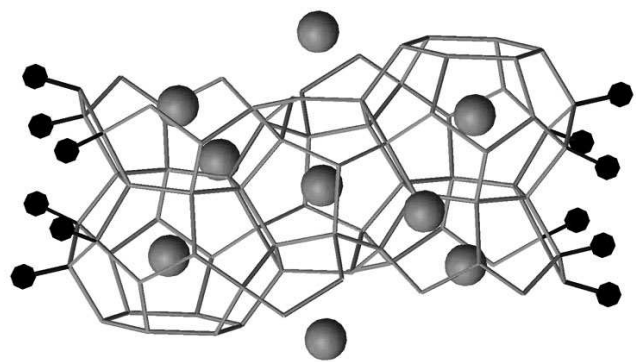}
\end{center}
\caption{ Symbolic representation of the Li$_8$@C-46 (100):H slab
unit cell. The structure consists of C$_{20}$ and C$_{24}$
clusters with carbon atoms (not shown here) located at the corners
of the cages. The gray balls show Li atoms encapsulated inside the
carbon clusters. The black balls represent hydrogen atoms on the
surface of the slab. The same type of slab was used for C-46 (100)
and C-46 (100):H surfaces.} \label{slab}
\end{figure}

We present in this Letter an {\it
ab-initio} study of the electronic properties of C-46 and
Li$_8$@C-46 within the density functional theory (DFT)
\cite{Hohen-Kohn-Sham}. We study in  particular the  C-46 (100)
bare and hydrogenated surfaces and analyze the effect of doping on
the position of the Fermi level, the affinity of the system and
the density of states  at E$_F$. We show that under doping
such a class of materials
are true metals with a low work function, thus combining
the advantages of diamond-like systems with a large carrier
concentration.

The local-density approximation (LDA) \cite{CA-PZ}
was used for all calculations  and
a pseudopotential approach was adopted \cite{TM-KB}. We used the
{\sc Siesta} package \cite{siesta}, which is a self-consistent
DFT code, employing numerical atomic-like orbitals (NAO) as a
basis set. A well-converged basis set, consisting of doubled
{\it \{s,p$_x$,p$_y$,p$_z$\}} orbitals plus polarization {\it
d}-orbitals was applied. Some of the calculations were also
performed using a standard plane-wave (PW) basis set.
All systems studied below were fully relaxed, both with
respect to cell size and atomic positions. To perform the
electronic affinity analysis, we calculated the plane-averaged
self-consistent potential in the direction perpendicular to the
surface for each of the slabs. The positions of the valence-band
maximum (VBM) and conduction-band minimum (CBM) were found by
adding the energy difference between the average self-consistent
potential in the bulk and the VBM (CBM) in the bulk to the average
self-consistent potential inside the slab.

\begin{table}
\caption{Calculated  electronic affinity for the diamond and
clathrate carbon surfaces (eV).}
\begin{ruledtabular}
\begin{tabular}{lccr}
{surface} & LDA-PWs & LDA-NAOs & GW    \\
\hline \\
C (100)          &  2.17   &  1.99   &  1.3   \\
C (100):H        & -0.45   & -0.59   & -1.3   \\
C-46 (100)       &         &  3.47   &  2.8   \\
C-46 (100):H     &         &  1.47   &  0.8   \\
Li$_8$@C-46 (100):H &      &  2.18   & $\sim$ 1.5   \\
\end{tabular}
\end{ruledtabular}
\label{table}
\end{table}


Before studying the C-46 clathrate phase, we have tested the method by
exploring the electronic and structural properties for the well-known
bare and hydrogenated diamond (100) surfaces. Following
Ref.\onlinecite{Weide94}, we have adopted a slab geometry with
10-layers but with 2-atoms per layer. The $\bf
k$-point sampling was thus increased  to a 4$\times$4$\times$1 special
grid \cite{MonkPack}. A vacuum of 10 \AA~was kept between
neighboring slabs.

The results for PW and NAO-basis are reported in Table
\ref{table}. As in Ref.\onlinecite{Weide94}, we observe that the
hydrogenation, by building a surface dipole, leads to a negative
affinity system. Our results are in good agreement with those
previously published \cite{Weide94}, and the NAO-basis
calculations reproduce within $\sim$ 0.15 eV  the results of
the PW calculations.


As the DFT-LDA is well known to underestimate the band gap of
insulators, we have performed further a quasiparticle study within
the GW approximation \cite{Hybertsen} for the bulk diamond and
C-46 phases. Such an approach is known to yield quasiparticle
energies to within 0.1 eV as compared to experimental
photoemission data. More details about the formalism and results
will be provided elsewhere \cite{xblase}. The main outcome of such
calculations is that the quasiparticle correction to the DFT-LDA
eigenstates are very similar in both systems and that the CBM is
pushed to higher energy by $\sim$ 0.7 eV in both phases (while the
VBM is corrected down by $\sim$ 0.8 eV).  This allows to correct
the calculated affinity as reported in the right column of Table
\ref{table}.

\begin{figure}
\begin{center}
\includegraphics[width=8.5cm]{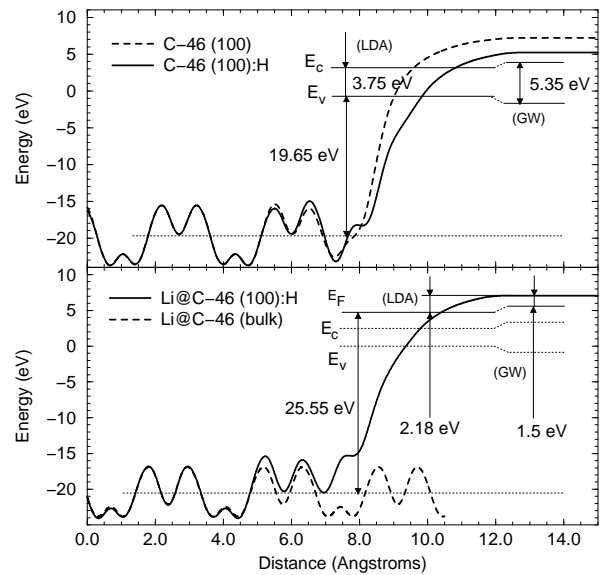}
\caption{ Plot of the plane-averaged self-consistent potential for
the C-46(100) bare and hydrogenated surfaces, and for
Li$_8$@C-46(100):H surface. The good agreement between the bulk
and slab potential on the innermost atoms of Li$_8$@C-46(100):H
indicates that the slab is large enough.} \label{potential}
\end{center}
\end{figure}


The stability and electronic properties of C-34 and C-46
clathrates have already  been discussed in some details at the
DFT-LDA level  \cite{Adams94,Nesper93,SanMiguel99,Saito97}. In
particular, it has been shown that for the C-34 and C-46
clathrates, the cohesive energy, bulk modulus and band gap are
respectively $\sim$ 0.1 eV/atom, $\sim$ 10 $\%$ and $\sim$ 0.2 eV
smaller than those of the  diamond phase. As such, the stability
and mechanical properties of carbon clathrates are close to those
of the diamond phase. In the present work, we obtained for the
lattice parameter, bulk modulus and band gap the values of 6.64
\AA, 376 GPa and 3.75 eV respectively with both NAO and PW basis
sets, which is in good agreement with previously reported results.
Within the GW approximation, the gaps increase to 5.55 and 5.35 eV
for the diamond and clathrate phases respectively.

We now study the C-46 (100) surface represented in Fig.\ref{slab}.
There are several non-equivalent (100) surfaces  and we adopt
a surface consisting of "half emerging" C$_{24}$ clusters.  After structural
relaxation, we find that the emerging semi-clusters hardly reconstruct
besides a slight contraction inward of the three-fold coordinated
surface atoms, leading to a minimum C-C distance of 1.40 \AA~ on
the surface. Hydrogen coverage is performed by "passivating" these
surface three-coordinated atoms. We note at this point that the results
presented below may certainly change quantitatively  with the chosen
surface but the purpose of the present paper is to illustrate  on
this specific example the potentialities of such novel systems.
Fig.\ref{potential} shows the calculated plane-averaged,
self-consistent potential for bare and hydrogenated C-46 (100)
surfaces. The electronic affinity is reported in Table
\ref{table}.

\begin{figure}
\begin{center}
\includegraphics[width=8cm]{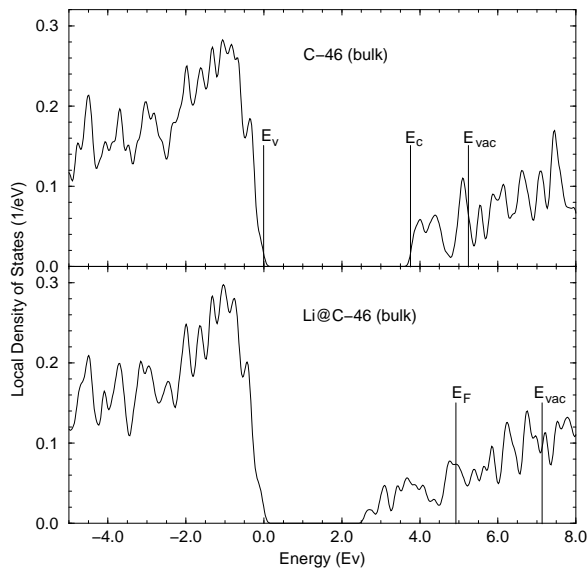} \caption{ Electronic density of states (eDOS) for
bulk C-46 and Li$_8$@C-46. A 0.1 eV broadening has been used. The
eDOS is normalized per carbon atom.} \label{dos}
\end{center}
\end{figure}

The results obtained for the undoped compounds suggest that carbon
clathrates are {\it a priori} less favorable for emission than,
say, the (100) diamond surface, taking the electronic affinity as
the only relevant criteria. However, we emphasize again that
clathrates, contrary to the diamond phase, can be efficiently
doped \cite{Saito97,Bernasconi00} by intercalating one atom in each cage
in order to obtain a true metallic system with a large density of
electrons available for emission. This is a crucial advantage as
compared to the diamond phase.

By analogy with the existing doped Si clathrates and intercalated
carbon graphitic systems, we study the {\it n}-doped Li$_8$@C-46 phase.
The corresponding bulk eDOS is represented in Fig.\ref{dos} where
we have arbitrarily  aligned the eDOS to the top of the C-46
valence bands for sake of comparison.
As can be readily seen, intercalation of Li atoms
leads to the filling of the conduction bands of the host carbon
matrix by Li 2s electrons. The system becomes metallic with the
Fermi level located near a peak in the density of states, which is in
excellent agreement with previous work \cite{Saito97}.

As in the case of diamond surfaces, hydrogenation leads to a significant
decrease in the electronic affinity of the empty phase.
We now explore the effect of doping on the electronic affinity
(Table \ref{table}) of the hydrogenated surface.
The important finding is that Li doping leads to
the formation of a true metallic system with a quite low
work function ($\sim$ 1.5 eV GW value).
This work function is significantly lower
than the one of alkali metals and much
lower than the one of e.g. tungsten. The eDOS for
states located around  E$_F$ can be further shown to have a strong
weight on  both bulk and surface atoms,
a factor which is crucial for emission.
One can conclude that doped carbon clathrates are potentially
true metals with a work function lower than the
lowest affinity of all elemental metallic systems but with much
superior mechanical properties.

It is interesting to note that the work function of the Li-doped
clathrate is actually larger than the electronic affinity of the
empty phase, despite the raise of the Fermi level under doping.
This means that the entire band structure
is shifted to lower energy with respect to the vacuum level under
doping. This shift can originate in two effects: a
decrease of the average potential inside the slab (bulk effect)
and an increase of the surface dipole barrier (surface effect).
An analysis of both effects show that in the present case it is
mainly a bulk effect.


In conclusion, we have shown that doped carbon clathrates  are
potentially true metallic systems with a work function significantly
lower than the one of alkali metals but with a stability
typical of the carbon diamond phase. This opens new perspectives
for the making  of efficient electron-emitters. An alternative choice
of dopant atom and/or surface treatment (e.g. cesium coverage) may
lead to even smaller work function.

\acknowledgments Computations  have been performed on the
supercomputer facilities (IDRIS) at the French "Centre National
de Recherche Scientifique" (CNRS). One of the
authors (VT) acknowledges a ``R\'egion Rh\^one-Alpes"
Postdoctoral Fellowship. The authors are indebted to Stephen Purcell
and C. Constancias (OpsiTech, Grenoble, France) for stimulating discussions.



\begin{references}
\bibitem{xblase}
In the case of the Li$_8$@C-46, the quasiparticle correction to E$_F$ is taken
to be the same as the one for the CBM of C-46. While in the conduction bands
the GW correction can be shown to slightly increase with the energy of the
states (see Ref.~\onlinecite{Hybertsen}), the enhanced screening
under doping is expected to reduce it. Despite such a compensation, the
quasiparticle correction to E$_F$ should be taken as an approximate value.
X. Blase, to be published.
\end{references}
\end{document}